\documentclass[aps,prresearch,floats,twocolumn,noshowpacs,superscriptaddress]{revtex4-2}

\usepackage{graphicx,epsfig}
\usepackage{times}
\usepackage{lipsum} 
\usepackage{amssymb,amsmath,multirow,rotate,color,mathtools}
\usepackage[utf8]{inputenc} 
\usepackage[dvipsnames]{xcolor}
\usepackage{float}
\usepackage{cancel}
\usepackage[colorlinks=true,
            linkcolor=blue,
            citecolor=blue,
            urlcolor=blue]{hyperref}

\begin{document}
\title{Species differences dilute the stabilizing effect of higher-order interactions in competitive communities}

\author{Marc Duran-Sala}
\affiliation{
Laboratory of Urban and Environmental Systems, École Polytechnique Fédérale de Lausanne (EPFL), Lausanne, Switzerland}

\author{Sandro Meloni}
\email[Corresponding author: ]{sandro@ifisc.uib-csic.es}
\affiliation{Institute for Cross-Disciplinary Physics and Complex Systems (IFISC), CSIC-UIB, Palma de Mallorca, Spain}
\affiliation{Centro Studi e Ricerche ``Enrico Fermi'' (CREF), Rome, Italy}

\author{Violeta Calleja-Solanas}
\affiliation{Department of Biology, University of Oxford, UK}
\affiliation{Estación Biológica de Do\~nana (CSIC), Seville, Spain}

\date{\today}

\begin{abstract}
Higher-order interactions, in which the presence of a third species modulates the interaction between two others, have been proposed as a mechanism that stabilizes ecological communities. In particular, in models of intransitive competition, replacing pairwise contests by three-way contests turns neutral oscillations into a stable coexistence equilibrium. Whether this stabilizing effect survives in more realistic scenarios, when pairwise and higher-order interactions act together, species differ in their demographic rates, and interactions take place on a network remains unclear.
Here, we study a competitive community model with a tunable fraction $\alpha$ of higher-order interactions, combining Lyapunov and linear stability analysis with numerical simulations and empirical birth and mortality rates of plants.
For identical species, any $\alpha>0$ makes the coexistence equilibrium globally stable. Differences among species in the ratio of birth to mortality rates, instead, make pairwise competition actively destabilizing and shift the transition to a finite critical fraction $\alpha_c$, which grows with the variance of these ratios. For larger differences, a feasible equilibrium often ceases to exist and, when it exists, higher-order interactions frequently fail to stabilize it. Even when stable, the coexistence equilibrium is often only one of two attractors near the transition, so that robust coexistence requires a larger fraction of higher-order interactions. On sparse random hypergraphs, a finite fraction is needed even for identical species: coexistence requires each individual to take part, on average, in more than one higher-order interaction, well beyond the percolation of hyperlinks. Our results show that the claimed stabilizing effect of higher-order interactions is fragile, and depends jointly on species differences and on the structure of interactions.
\end{abstract}

\maketitle

\section{Introduction \label{sec:1}}

The question of how different species coexist has captivated researchers beyond ecology, including disciplines like statistical physics and mathematics. Challenges mainly arise from the competitive exclusion principle, which asserts that inferior competitor species should eventually be driven to extinction by better-adapted counterparts. Among other coexistence mechanisms, intransitive competition has gained attention as a potential bypass of exclusion \cite{soliveres_everything_2018}. Intransitive competition establishes no hierarchy among species, such as in a ``rock-paper-scissors'' tournament \cite{kerr_local_2002}. 
However, models implementing intransitive dominance result in abundances neutrally cycling around an equilibrium point, or asymptotic solutions in which the system cycles from being composed almost entirely of one species to almost wholly by another and so forth \cite{may_nonlinear_1975}; something that is unlikely to occur in nature. In addition, the presence of large oscillations may turn against coexistence since species could become extinct by external perturbations or stochastic events. Moreover, evidence of systems that rely completely on intransitive competition as the main mechanism for promoting coexistence remains scarce \cite{soliveres_everything_2018, godoy_intransitivity_2017}. 

For these reasons, theoretical studies aim to overcome the large oscillations by finding conditions for stable coexistence \cite{grilli_higher-order_2017} (i.e., local asymptotic stability of a feasible coexistence equilibrium \cite{may_will_1972}) and explain when intransitivity could be found in nature as a coexistence mechanism \cite{pajaresmurgo_intransitivity_2024,calleja2025general}. Their approach is to combine intransitivity with auxiliary mechanisms, such as mobility \cite{reichenbach_mobility_2007} or spatial interactions \cite{calleja-solanas_structured_2022,dieckmann_geometry_2000,lowery_structured_2019}. 

In this context, Higher-Order Interactions (HOIs) have been proposed as another mechanism to stabilize ecological communities, e.g.~\cite{grilli_higher-order_2017}. HOIs emerge when the presence of a third species influences the interaction between the other two~\cite{Golubski2016,levine_beyond_2017}. Multispecies dependencies are, of course, not new in ecological theory: consumer--resource models, saturating functional responses, interaction modifications in food webs, and competition for space all couple the dynamics of many species at once, and effective higher-order terms emerge naturally when such mechanistic models are reduced to phenomenological descriptions~\cite{letten_mechanistic_2019,terry_trophic_2017,kleinhesselink_detecting_2022}. Likewise, higher-order dynamics can arise on ordinary networks, and pairwise and higher-order representations of the same system are not always distinguishable~\cite{young_hypergraph_2021}. Here, we use the term HOIs in the narrower sense of explicit interaction terms involving three species, whose effect is not a sum of pairwise terms. They attracted much attention in ecological research several decades ago, in part because the intrinsic nonlinearity promised greater realism than the classic Lotka–Volterra equations \cite{abrams_arguments_1983,billick_higher_1994}. However, their popularity waned amid debates over their definition \cite{pomerantz_higher_1981,wootton_putting_1994} and detection \cite{case_testing_1981,yodzis_indeterminacy_1988}. 

In recent years, HOIs have experienced a revival of interest~\cite{levine_beyond_2017,battiston_networks_2020,battiston_physics_2021,gibbs2024can}. Although some concerns remain unresolved~\cite{kleinhesselink_detecting_2022, letten_mechanistic_2019}, a growing body of theoretical work indicates that HOIs can substantially influence community diversity and stability ~\cite{bairey_high-order_2016,gibbs_coexistence_2022,terry2025impact,Li2026,LechonAlonso2026tipping}. However, most existing models rely on idealized assumptions, such as species equivalence and well-mixed interactions among  individuals. Consequently, whether HOIs can stabilize competitive communities under more realistic conditions remains an open question.


Here, we address this  gap by studying a model with a variable fraction of higher-order interactions and identifying under which conditions stable coexistence emerges or breaks.  By using analytical derivations and numerical simulations with synthetic and empirical physiological rates of plants, we explore how the critical fraction of HOIs depends on the heterogeneity of physiological rates, both in well-mixed communities and on random interaction networks.
As a starting point, we test the influence of HOIs among species with equal physiological rates by varying their proportion in situations where pairwise competition alone cannot stabilize the system. Then, we explore how variability in physiological rates affects the critical proportion of HOIs needed to stabilize the dynamics. Finally, we abandon the well-mixed scenario and define an interaction network, whose nodes are single individuals of different species connected by links and hyperlinks, representing pairwise and higher-order interactions respectively\cite{battiston_networks_2020}. 


Our results challenge the claimed stabilizing effect of HOIs. While for equal physiological rates even a minimal presence of HOIs leads to stable coexistence, heterogeneity in the rates makes pairwise competition destabilizing and dilutes the stabilizing effect of HOIs in the well-mixed scenario. Moreover, when individuals interact through a sparse random network, stability requires a finite fraction of HOIs even for identical species. 



\section{Model \label{sec:2}}


\begin{figure*}[t]
  \includegraphics[width=\linewidth]{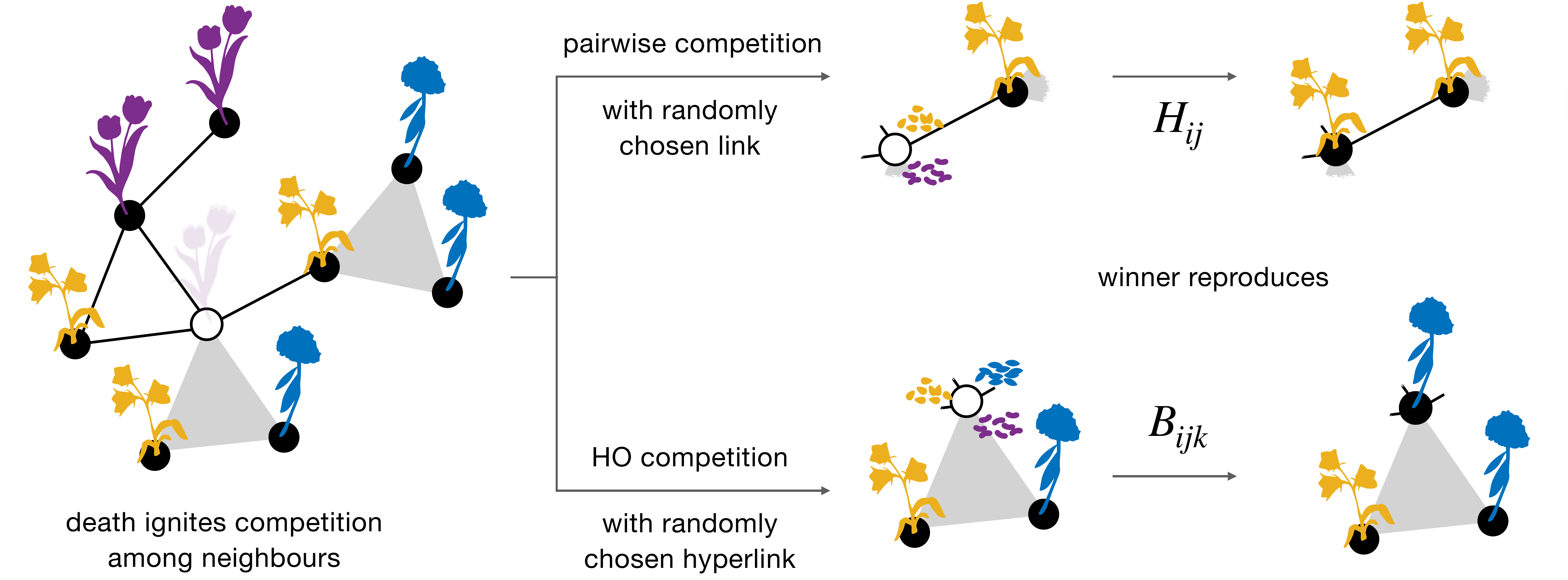}
  \caption{\textbf{Illustration of the competitive dynamics on an interaction network}. When a random plant dies, it leaves a vacant site, where seeds of its own species remain. One of the links or hyperlinks of the vacant site is chosen at random, and the seeds of the dead plant's species compete with those of the plants connected by that link (pairwise competition, top) or hyperlink (higher-order competition, bottom). The winner, chosen according to the dominance matrix $H$ or the tensor $B$, respectively, reproduces and occupies the site.}
\label{fig:1}
\end{figure*}

\subsection{Competitive community model with higher-order interactions \label{sec:2.1}}
We consider an isolated competitive community formed by $g$-species. To focus only on the interplay between pairwise and higher-order interactions (HOIs), we keep the number of processes to a minimum. Each species $i$ has two physiological rates, specifically birth $f_i$ and mortality $d_i$ rates, and competes with the other species.

Figure~\ref{fig:1} provides an illustration of the model in the context of flowering plants competing for space. At each iteration, a plant dies, creating an opening. One link of the vacant site is selected at random, and the seeds left by the dead plant compete with those of the plants that belong to that link to establish a seedling there (Fig.~\ref{fig:1}). Depending on the link order, the winner is determined either by the $g\times g$ dominance matrix $H$ for pairwise interactions or by the $g\times g\times g$ dominance tensor $B$ in the case of HOIs. In a well-mixed population, this process leads to the mean-field equations below (see Appendix~\ref{sec:methods_mf} for the derivation).

Mathematically, the evolution of species density $x_i$, with $\sum_i x_i(t)=1$ at every $t$, can be described by replicator dynamics. In the pairwise case it reads as:
\begin{equation}
\begin{aligned}
\frac{dx_i}{dt} =  x_{i} \left(  \frac{D(x)}{F(x)^{2}} f_{i} \sum_{j}  2 H_{i j} f_{j} x_{j}-d_{i}\right),
\end{aligned}
\label{eq:pair}
\end{equation}
where the dominance matrix $H$ encodes the competitive interactions among species. In particular, $H_{ij}$ is the probability species $i$ outcompetes species $j$, so that $H_{ii} = 0.5$ and $H_{ij} + H_{ji} = 1$. Introducing a probabilistic dominance matrix allows us to go beyond neutral competition $(H_{ij} = 0.5 \ \forall i,j)$ and complete $(H_{ij} = 1, H_{j,i} = 0 \ \forall i,j)$ dominance between species \cite{grilli_higher-order_2017}. The developing probability of offspring of species $i$ is $f_i x_i /F(x)$, given that $F(x) = \sum_i f_i x_i$, and $d_i x_i /D(x)$ is the dying probability of an individual belonging to species $i$, where $D(x)= \sum_i d_i x_i$. Varying these physiological rates also allows us to break neutrality in the classic sense of species with equal probabilities of reproduction and death \cite{leroi_neutral_2020}.

Conversely, one could also express the dynamics of a system of purely higher-order interactions. If we limit ourselves to \textit{triplewise} interactions, the equations are
\begin{equation}
    \frac{dx_i}{dt} = x_{i} \Bigg(\frac{ D(x)}{F(x)^{3}} f_{i} \sum_{j,k } B_{ijk} f_{j} x_{j} f_{k} x_{k}-d_{i} \Bigg),
    \label{eq:ho}
\end{equation}
in which $B$ is a tensor (in the sense of a multidimensional array) whose elements $B_{ijk}=B_{ikj}$ weight the success of species $i$ in a three-way contest with species $j$ and $k$. In the following, we assume, as in~\cite{grilli_higher-order_2017}, that the outcome of a three-way contest is decided by a tournament of pairwise encounters, so that  $B_{ijk}=2H_{ij}H_{ik} + H_{ij}H_{jk} + H_{ik}H_{kj}$. With this choice $B_{ijk}+B_{jik}+B_{kij}=3$ for any dominance matrix $H$, so that $B_{ijk}\in[0,3]$ and $B_{ijk}/3$ is the probability that species $i$ wins the three-way contest. This normalization guarantees $\sum_i dx_i/dt=0$.

We stress that the term ``higher-order'' can refer to three distinct notions: (i) the topology of the interactions, i.e. hyperlinks connecting more than two individuals; (ii) terms of order higher than two in the dynamical equations; and (iii) genuine ecological higher-order effects, in which a third species modifies the per-capita interaction between two others in a way that is not reducible to pairwise outcomes~\cite{levine_beyond_2017}. Our model implements (i) and (ii), while, because $B$ is built from $H$, it represents a specific, tournament-type class of (iii). 

To model the presence of both pairwise and higher-order interactions at the same time, we introduce a parameter $\alpha \in [0,1]$ that controls their proportion: $\alpha$ is the probability that a contest for a vacant site is a three-way rather than a pairwise one. Thus, the full model becomes:
\begin{equation}
\begin{aligned}   
    \frac{dx_i}{dt} &= (1-\alpha) \left[ x_{i} \left(  \frac{D(x)}{F(x)^{2}} f_{i} \sum_{j}  2 H_{i j} f_{j} x_{j}-d_{i}\right) \right] \\
    & + \alpha \left[
    x_{i} \Bigg(\frac{ D(x)}{F(x)^{3}} f_{i} \sum_{j,k } B_{ijk} f_{j} x_{j} f_{k} x_{k}-d_{i} \Bigg)\right].
\end{aligned}
\label{eq:general}
\end{equation}
Equations~\eqref{eq:pair} and~\eqref{eq:ho} define the two limiting cases of our framework, with $\alpha=0$ corresponding to purely pairwise competitive interactions and $\alpha=1$ to purely higher-order interactions. Although simplified, our model is quite general and can be mapped onto an effective generalized Lotka–Volterra (GLV) dynamics~\cite{HOFBAUER1981,Hofbauer_Sigmund_1998,allesina2026equivalence}. In the $\alpha=0$ limit with equal physiological rates, species densities neutrally oscillate if the dominance matrix has intransitive cycles \cite{kerr_local_2002,allesina_competitive_2011}. This limit is structurally unstable, and we use it only as a reference point: as we show below, with heterogeneous physiological rates the pairwise dynamics is no longer neutral but actively destabilizing, and the transition to stability becomes generic. At the other end, when physiological rates are equal or very similar, the $\alpha=1$ limit recovers the stabilizing scenario studied in \cite{grilli_higher-order_2017}, leading to a stable fixed point. We therefore focus on the transition between these two limiting cases, where $\alpha_c$ represents the critical fraction of higher-order interactions needed for stabilizing the dynamics.

\begin{figure*}[t]
    \centering
    \includegraphics[width=\linewidth]{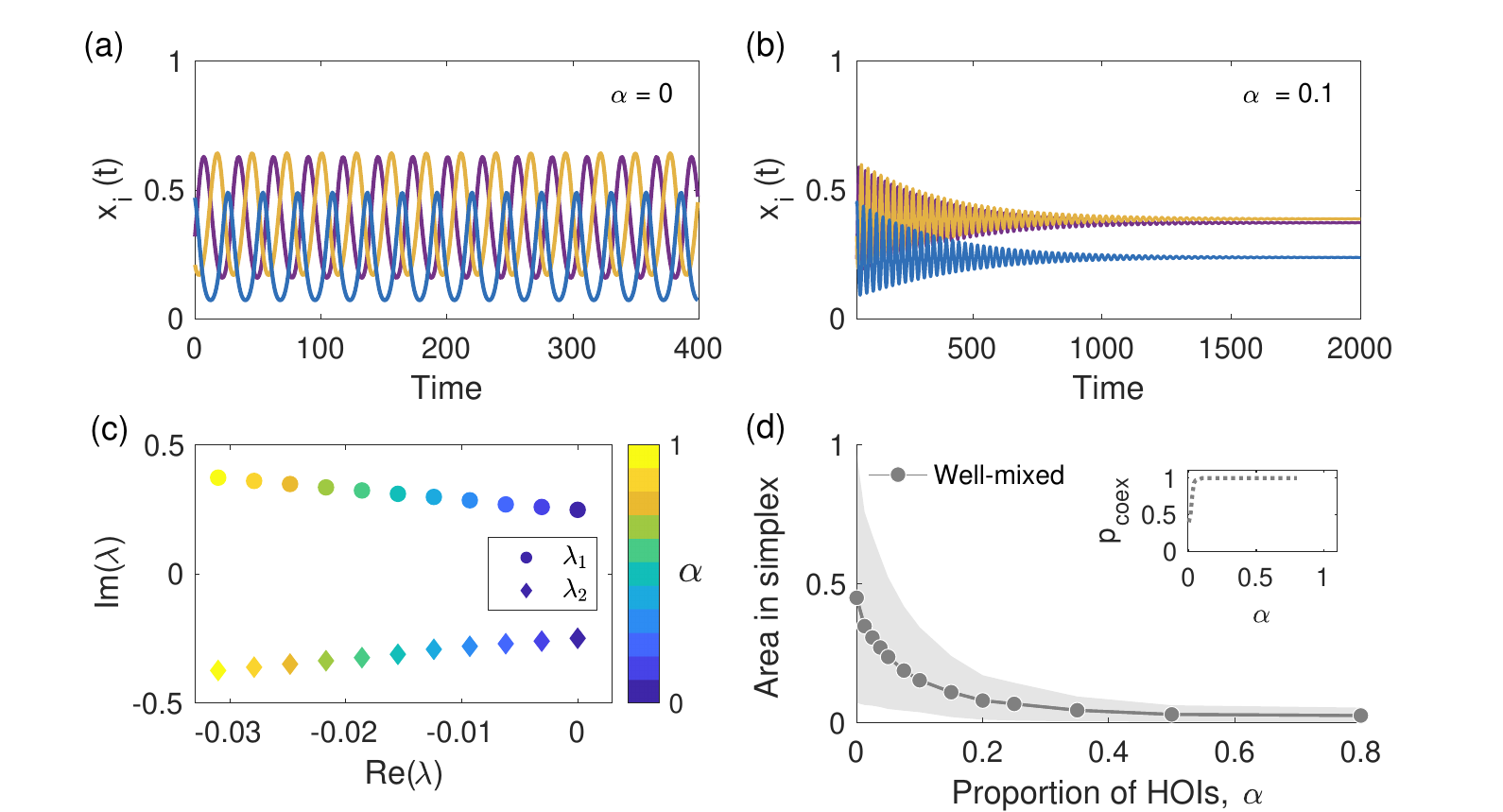} 
    \caption{\textbf{Any proportion of higher-order interactions stabilizes the dynamics in well-mixed competitive systems with equal physiological rates.} Numerical integration of the dynamics from the set of ODEs, Eq.~\eqref{eq:equal}, for (a) $\alpha = 0$ (no HOIs) and (b) $\alpha=0.1$. (c) Eigenvalues for different proportions of HOIs (dots color) at the fixed point. For $\alpha > 0$, the pair of complex conjugate eigenvalues cross into the left half-plane, guaranteeing that the fixed point is stable. (d) Average area on the simplex of our $3$-species system as a function of different values of $\alpha$. The areas are calculated over $50$ Monte Carlo simulations. Shades indicate the standard deviation. The inset is the probability of coexistence $p_{coex}$ over $100$ simulations since the noisy nature of the simulations may lead to extinctions when the oscillations are wide enough. To measure the area in the simplex, we only consider the dynamics of those systems in which all species coexist. All panels are computed using the dominance matrix introduced in \cite{calleja-solanas_structured_2022}.}
    \label{fig:2}
\end{figure*}


Although the analytical framework is not restricted to three species, in the following we focus on the case of $g=3$  because it represents the minimal setting under which HOIs stabilize the dynamics.

Finally, to characterize the stability of the system during Monte Carlo simulations, we do not consider directly the amplitude of the oscillations in $x_i(t)$.  Due to the stochastic nature of the simulations, it could be difficult to disentangle stochastic noise from limit cycle oscillations, possibly leading to misleading results. Instead, we take advantage of the fact that species densities can be interpreted as a point tracing a trajectory within a $g-1$ simplex covering the space of possible ecological states, where each vertex corresponds to a single-species population. As time evolves, the area encircled by the densities’ trajectories on the simplex can be used to characterize the macroscopic state of the system \cite{calleja-solanas_structured_2022} (see Sec.~S1 of the Supplemental Material (SM)~\cite{SM} for details). The trajectory occupies a small area when the system fluctuates with low amplitude around a certain point. In contrast, larger oscillations cover a wider area in the simplex.

\section{Results \label{sec:3}}

\subsection{For well-mixed systems with equal physiological rates, any proportion of HOIs is sufficient for stability  \label{sec:3.1}}


We begin our analysis by studying the simplified case of equal physiological rates ($d_i=f_i=1  \  \forall i$), for which Eq.~\eqref{eq:general} reads:

\begin{equation}
\begin{aligned}
   \frac{dx_i}{dt} = (1 - \alpha) \Bigg( -x_i + 2 \sum_jH_{ij} x_i x_j \Bigg) \\
   + \alpha \Bigg( - x_i + \sum_{j,k} B_{ijk} x_i x_j x_k \Bigg) .
\end{aligned}
\label{eq:equal}    
\end{equation}\\

The equilibrium of Eq.~\eqref{eq:pair} $(2\sum_j H_{ij}x_j^* = 1)$  is also an equilibrium of this equation for every $\alpha$, since $\sum_{j,k}B_{ijk}x^*_jx^*_k=1$ whenever $2\sum_j H_{ij}x_j^*=1$. We assume it to be feasible ($x^*_i > 0 , \  \forall i$ \cite{saavedra2017structural}) and characterize its stability in two complementary ways: globally, through a Lyapunov function, and locally, through the Jacobian matrix. For the global analysis we use the function (see \cite{grilli_higher-order_2017})
\begin{equation}
V(x)=-\sum_{i} x_{i}^{*} \log \frac{x_{i}}{x_{i}^{*}},
\label{eq:Lyapunov}
\end{equation}
\noindent
which is non-negative on the simplex and vanishes only at $\Vec{x}^*$. Differentiating Eq.~\eqref{eq:Lyapunov} along the trajectories of Eq.~\eqref{eq:equal}, we obtain
\begin{equation}
    \begin{aligned}
        \frac{dV}{dt} &=\sum_{i} \frac{\partial V}{\partial x_{i}} \frac{\mathrm{d} x_{i}}{\mathrm{~d} t} =-\sum_{i} \frac{x_{i}^{*}}{x_{i}} \frac{\mathrm{d} x_{i}}{\mathrm{~d} t} \\
        &=\alpha \left[ -2 \sum_{i} x_{i}^{*}\left(\sum_{j} H_{i j} \xi_{j}\right)^2 \right],
    \end{aligned}
    \label{eq:Lyapunov_derivation1}
\end{equation}
where we have introduced $\xi_j \coloneqq x_j - x^*_j$. For $\alpha=0$, we recover the case of only pairwise interactions, where $dV/dt = 0$: $V$ is a constant of motion and the system follows neutral cycles around the equilibrium (Fig.~\ref{fig:2}a). When $\alpha>0$, $dV/dt \leq 0$, with equality only at $\Vec{\xi}=0$ when $H$ is full rank. By LaSalle's invariance principle, $\Vec{x}^*$ is then the unique and globally asymptotically stable fixed point in the interior of the simplex (Fig.~\ref{fig:2}b).

The local picture is equally simple. Since $\Vec{x}^*$ does not depend on $\alpha$, the Jacobian of Eq.~\eqref{eq:equal} at the equilibrium, restricted to the simplex (the reduced Jacobian $J$ of Appendix~\ref{sec:methods_jac}), is linear in $\alpha$: $J(\alpha)=(1-\alpha)J_{\mathrm{p}}+\alpha J_{\mathrm{h}}$, where $J_{\mathrm{p}}$ and $J_{\mathrm{h}}$ are the reduced Jacobians of the purely pairwise and purely higher-order dynamics. For $g=3$, $J$ is a $2\times2$ matrix with eigenvalues $\lambda_\pm=\left(\tau\pm\sqrt{\tau^2-4\Delta}\right)/2$, where $\tau=\mathrm{tr}\,J$ and $\Delta=\det J$. At $\alpha=0$ the equilibrium is a center, $\tau(0)=0$ and $\Delta(0)>0$, while $\mathrm{tr}\,J_{\mathrm{h}}<0$ as a consequence of Eq.~\eqref{eq:Lyapunov_derivation1}. Hence $\mathrm{Re}\,\lambda_\pm=\alpha\,\mathrm{tr}\,J_{\mathrm{h}}/2$, and this holds for every $\alpha$, since the eigenvalues remain complex conjugate [their exact expression is given in Eq.~\eqref{eq:jac_eig}]: the pair of complex-conjugate eigenvalues crosses the imaginary axis transversally at $\alpha_c=0$ (Fig.~\ref{fig:2}c). Note that, because the $\alpha=0$ system is conservative, this is a degenerate Hopf point: no isolated limit cycle is born, and the transition is from a family of neutral cycles to a stable focus. In summary, with equal physiological rates any nonzero fraction of HOIs makes the coexistence equilibrium both locally and globally stable.

\begin{figure*}[t]
    \centering
    \includegraphics[width=0.9\linewidth]{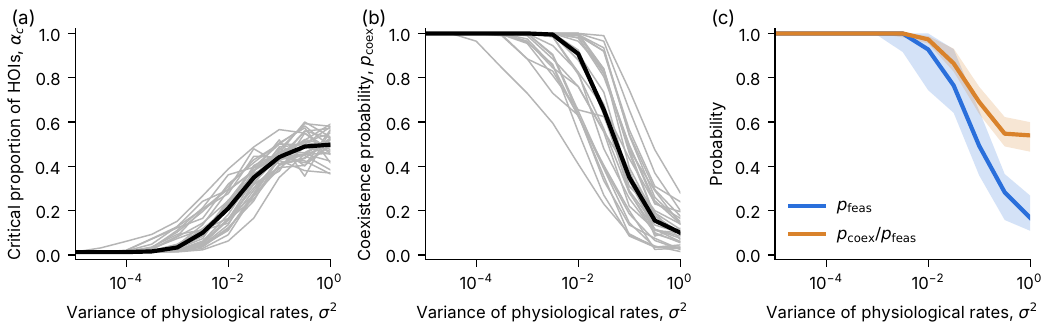}
    \caption{ \textbf{As variability in physiological rates increases, a greater proportion of higher-order interactions is needed to stabilize the dynamics.} Linear stability analysis of Eq.~\eqref{eq:general} with physiological rates sampled from $\mathcal{N}(1,\sigma^2)$, for $30$ intransitive dominance matrices $H$ and $40$ triplets of species per value of $\sigma^2$; $\alpha$ is scanned on a grid of step $0.0125$, which sets the lower limit of $\alpha_c$ in panel (a). (a) Critical proportion of higher-order interactions $\alpha_c$, averaged over the triplets for which it exists. (b) Probability $p_{\textrm{coex}}$ that a feasible and stable equilibrium exists for some $\alpha$. In (a) and (b), grey lines correspond to individual dominance matrices $H$ and the black line is the median across $H$. (c) Probability $p_{\textrm{feas}}$ that a feasible equilibrium exists for some $\alpha$ (blue), and conditional probability $p_{\textrm{coex}}/p_{\textrm{feas}}$ that HOIs can stabilize it (orange). Lines are medians and shaded regions the interquartile range across $H$.}
    \label{fig:3}
\end{figure*}

\begin{figure*}[t]
    \centering
    \includegraphics[width=0.9\linewidth]{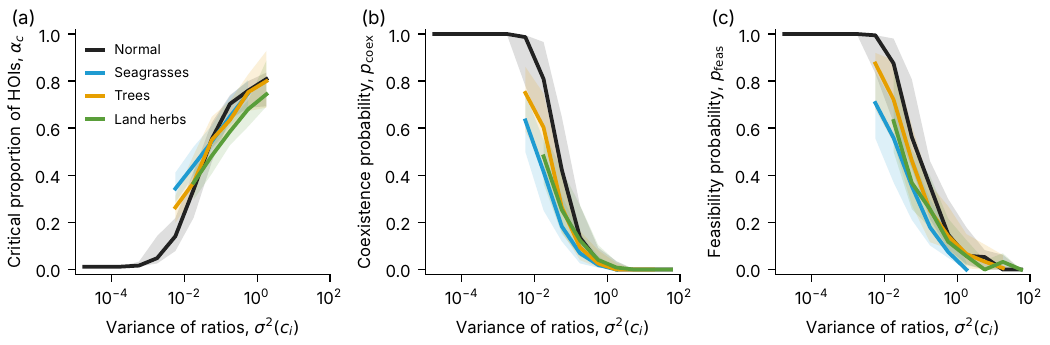}
    \caption{ \textbf{The effect of variability in physiological rates on stability persists under empirical distributions.} Linear stability analysis of Eq.~\eqref{eq:general} with empirical physiological rates from \cite{marba_allometric_2007} for three categories of plants ($150$ triplets per category and dominance matrix, $30$ dominance matrices $H$), compared with normally distributed rates (data of Fig.~\ref{fig:3}). Results are binned by the variance of the ratios of physiological rates $\sigma^2(c_i)$ within each triplet; bins with fewer than $8$ triplets are omitted. (a) Critical proportion of HOIs $\alpha_c$, averaged over triplets for which it exists. (b) Probability $p_{\textrm{coex}}$ that a feasible and stable equilibrium exists for some $\alpha$. (c) Probability $p_{\textrm{feas}}$ that a feasible equilibrium exists for some $\alpha$. Solid lines are medians and shaded regions the interquartile range across $H$. Note that the x-axis differs from that of Fig.~\ref{fig:3}.}
    \label{fig:4}
\end{figure*}
To corroborate these analytical predictions, we run Monte Carlo simulations of our system with well-mixed population of $g=3$ species and random initial conditions. During the simulations, extinctions can easily occur for $\alpha=0$ due to stochastic fluctuations (see inset in Fig.~\ref{fig:2}d). When $\alpha > 0$, the system stabilizes around the predicted equilibrium, which agrees with our theoretical derivations. In the simulations, the transition between these two regimes occurs smoothly as $\alpha$ increases. Nevertheless, the area covered by the trajectory decreases, and by $\alpha=0.1$, the system could be considered to have reached a fixed point (Fig.~\ref{fig:2}d). Hence, we do not need to lose pairwise interactions altogether to stabilize the dynamics: any proportion of higher-order interactions is enough to obtain the same effect.

\subsection{Species differences make stability conditional on the fraction of HOIs \label{sec:3.2}}
Up to this point, we have considered only species with identical birth and mortality rates; consequently, species differ solely in their competitive abilities (encoded in $H$). We now go beyond this assumption and study Eq.~\eqref{eq:general} with different physiological rates. In this case the equilibrium is no longer shared by the pairwise and higher-order dynamics, and it varies with the proportion of HOIs, $\alpha$. Defining
\begin{equation}
\begin{split}
    &T_{1,i}(\Vec{x}) :=  \frac{D}{F^2}\, f_i \sum_j 2 H_{ij} f_j x_j  \quad \text{and}\\
    &T_{2,i}(\Vec{x}) :=  \frac{D}{F^3}\, f_i \sum_{j,k} B_{ijk} f_j f_k x_j x_k
\end{split}
    \label{eq:T}
\end{equation}
(where we have dropped the arguments of $D(x)$ and $F(x)$ for readability), a feasible equilibrium $\Vec{x}^*(\alpha)$ must satisfy, for every species $i$,
\begin{equation}
    (1-\alpha)\,T_{1,i}^* + \alpha\, T_{2,i}^* = d_i ,
    \label{eq:equilibrium_diff}
\end{equation}
where $T^*_{\cdot,i} \coloneqq T_{\cdot,i}(\Vec{x}^*)$. Note that a solution of Eq.~\eqref{eq:equilibrium_diff} with all $x^*_i>0$ is not guaranteed to exist: feasibility and stability must be assessed separately.

\paragraph*{Pairwise competition between different species is destabilizing.}
The Lyapunov function of Eq.~\eqref{eq:Lyapunov}, evaluated with respect to $\Vec{x}^*(\alpha)$, still provides useful information. Using the equilibrium condition, its time derivative can be written exactly as (see Appendix~\ref{sec:methods_lya})
\begin{equation}
    \frac{dV}{dt} = \frac{\left(F^*D - FD^*\right)^2}{F^2D^*}  + \alpha\, G(\Vec{x}),
    \label{eq:dV}
\end{equation}
where $F^*=F(\Vec{x}^*)$, $D^*=D(\Vec{x}^*)$, $T^*_{2,i}=T_{2,i}(\Vec{x}^*)$, and
\begin{equation}
    G(\Vec{x}) = \frac{2DF^*}{F} - \sum_i x^*_i T_{2,i} - \frac{F^{*2}D}{F^2D^*}\sum_i x_i T^*_{2,i}
    \label{eq:G}
\end{equation}
collects the contribution of HOIs. Two consequences follow. First, in the purely pairwise case ($\alpha=0$), $dV/dt\geq 0$: $V$ cannot decrease, so a feasible equilibrium can never be asymptotically stable, and, unless the first term vanishes identically, trajectories move away from it, typically towards the boundary of the simplex. The same conclusion follows locally from the trace of the reduced Jacobian, which at $\alpha=0$ is strictly positive whenever the ratios $f_i/d_i$ are not all equal [Eq.~\eqref{eq:jac_pairwise_trace}]. The first term vanishes identically if and only if the ratio $F/D$ is constant along trajectories, i.e. when birth rates are proportional to mortality rates with the same constant for all species, $f_i = c\, d_i\ \forall i$. In that case the $\alpha=0$ dynamics is again conservative and, as for equal rates, any $\alpha>0$ stabilizes the equilibrium (we verified this numerically for all sampled cases). Therefore, the relevant measure of species differences is not the spread of birth or mortality rates separately, but the spread of their ratios $c_i=f_i/d_i$ across species. Second, for $\alpha>0$, $G$ is not sign-definite, so Eq.~\eqref{eq:dV} does not by itself provide a stability threshold; we determine it below through linear stability analysis.


\paragraph*{Linear stability analysis.}
We locate numerically, for each value of $\alpha$, the feasible solutions of Eq.~\eqref{eq:equilibrium_diff} and compute at each of them the reduced Jacobian $J(\alpha)$ of Eq.~\eqref{eq:general}. For $g=3$, $J$ is a $2\times 2$ matrix and an equilibrium is locally asymptotically stable if and only if $\tau(\alpha)=\mathrm{tr}\,J<0$ and $\Delta(\alpha)=\det J>0$. We define the critical fraction of HOIs, $\alpha_c$, as the smallest $\alpha\in[0,1]$ for which a feasible and locally stable equilibrium exists. If no such $\alpha$ exists, the HOIs cannot stabilize coexistence for that set of species. For each set of species we also record whether a feasible equilibrium exists for at least one $\alpha\in[0,1]$, which allows us to separate the loss of feasibility from the loss of stability. Details of the numerical procedure, and a comparison with direct integration of Eq.~\eqref{eq:general}, are given in Appendix~\ref{sec:methods_jac}. In all the cases we examined, once stability is reached it persists for every larger $\alpha$, so that $\alpha_c$ is well defined.

\paragraph*{Synthetic physiological rates.}
We first sample birth and mortality rates from normal distributions $\mathcal{N}(1, \sigma^2)$ truncated at zero (negative values are discarded and resampled), so that $\sigma^2$ controls the level of species differences. Note that equal mean birth and mortality rates do not imply proportional rates for each species: for any $\sigma^2>0$ the ratios $c_i$ differ among species and, as discussed above, $\alpha_c>0$. We repeat the analysis across an ensemble of $30$ dominance matrices $H$ randomly generated under constraints ensuring intransitive competition (see SM Section~S2 for details). For each value of $\sigma^2$, we define $p_{\text{feas}}$ as the fraction of sampled triplets of species for which a feasible equilibrium exists for some $\alpha$, and $p_{\text{coex}}$ as the fraction for which $\alpha_c$ exists, i.e. for which a feasible and stable equilibrium exists for some $\alpha$. The ratio $p_{\text{coex}}/p_{\text{feas}}$ is thus the probability that HOIs can stabilize a feasible equilibrium.

Fig.~\ref{fig:3}a shows that, as the variance of physiological rates becomes larger, $\alpha_c$ increases too, reaching values close to $0.5$ across dominance matrices. In addition, for a given dominance matrix $H$, $\alpha_c$ becomes increasingly variable as the variability of physiological rates increases, hindering predictions of community stability (see SM Figs.~S2--S5). Moreover, above a critical value of the variance, the probability $p_{\text{coex}}$ that some fraction of HOIs stabilizes the dynamics decreases abruptly (Fig.~\ref{fig:3}b): for many sets of species, no fraction of HOIs is enough to stabilize coexistence. This decrease has two distinct origins (Fig.~\ref{fig:3}c). For large species differences, a feasible equilibrium often does not exist for any $\alpha$.
In addition, even when a feasible equilibrium exists, HOIs are increasingly unable to stabilize it: the conditional probability $p_{\text{coex}}/p_{\text{feas}}$ decreases from $1$ at small $\sigma^2$ to $0.75$ at $\sigma^2=0.1$ and $0.5$ at $\sigma^2=1$. Loss of feasibility is thus the dominant effect at large species differences, but loss of stability is substantial as well.

\paragraph*{Local stability is not sufficient for robust coexistence.}
Linear stability guarantees that small perturbations of the equilibrium decay, but not that the equilibrium attracts the whole simplex. To test this, we took one triplet of species with $\sigma^2=0.04$ for each of the $30$ dominance matrices and integrated Eq.~\eqref{eq:general} from $20$ random initial conditions (see Appendix~\ref{sec:methods_lya}). At $\alpha=\alpha_c$, all trajectories converged to the equilibrium in only $10$ of the $30$ cases, while in $13$ cases most trajectories instead approached the boundary of the simplex, where one species is driven to extinction. Close to the transition the stable equilibrium thus coexists with an attracting heteroclinic cycle along the boundary, and its basin of attraction is small. Increasing the fraction of HOIs to $\alpha_c+0.05$ enlarges the basin: all trajectories converged in $15$ cases and at least $19$ out of $20$ in $18$ cases, but in $5$ cases the majority of trajectories still reached the boundary. The value $\alpha_c$ obtained from linear stability is therefore a lower bound for the fraction of HOIs needed for robust coexistence.
Hence, coexistence is unlikely when the ratios of birth to mortality rates differ too much among species.

\subsection{Empirical physiological rates}
So far, the physiological rates we considered were based on synthetic data. However, a natural question arises: how do our results change in a more realistic setting where physiological rates are based on empirical values found in real-world communities? To answer this question, we now sample physiological rates from an empirical dataset from a study investigating allometric scaling in plants \cite{marba_allometric_2007}. Importantly, our goal is to use these empirical distributions as a proxy for realistic variability in physiological rates, and ask how such variability affects the stabilizing role of higher-order interactions within our theoretical framework. To be more specific, we consider data from three different plant types: \textit{Trees}, \textit{Land and salt marsh herbs} and \textit{Seagrasses}, with $58$, $53$, and $131$ estimates of birth and mortality rates (from $37$, $29$, and $21$ species), respectively, spanning more than four orders of magnitude. 
For each plant type and each of the $30$ dominance matrices, we sample $150$ triplets of estimates $(f_i,d_i)$ and quantify species differences through the variance $\sigma^2(c_i)$ of the ratios $c_i = f_i / d_i$ within the triplet. We then apply the same linear stability analysis as in the Gaussian sampling case to compute $\alpha_c$, $p_{\text{feas}}$ and $p_{\text{coex}}$.

Empirical triplets are characterized by large differences in the ratio of birth to mortality rates: the median of $\sigma^2(c_i)$ ranges from $0.2$ (seagrasses) to $0.7$ (land and salt-marsh herbs), and fewer than $5\%$ of the triplets have $\sigma^2(c_i)<0.01$, the range where HOIs stabilize nearly all synthetic communities. Consequently, coexistence is rare: a feasible equilibrium exists for some $\alpha$ in only $17$--$25\%$ of the triplets, and HOIs stabilize it in $9$--$15\%$ of them, with a median $\alpha_c$ between $0.48$ and $0.54$ when a transition exists. When compared at the same value of $\sigma^2(c_i)$, the critical fraction $\alpha_c$ of empirical triplets is similar to that of synthetic, normally distributed rates (Fig.~\ref{fig:4}a), whereas both $p_{\text{coex}}$ and $p_{\text{feas}}$ are lower (Fig.~\ref{fig:4}b,c): at $\sigma^2(c_i)\approx 0.06$, for instance, the median $p_{\text{coex}}$ is $0.20$--$0.26$ for empirical rates and $0.49$ for synthetic rates. Thus, empirical heterogeneity in physiological rates further reduces the ability of HOIs to stabilize coexistence, mainly by making a feasible equilibrium less likely. Results for individual dominance matrices are provided in the SM (Figs.~S2--S5).

\subsection{Interactions on random hypergraphs\label{sec:3.3}}

So far, we focused our analysis on the case of well-mixed populations. However, assuming that each individual interacts with the entire population could be unfeasible in most scenarios. To overcome this limitation, we focus on Erd\"os-R\'enyi (ER) networks, placing each individual on a node and letting them interact, both under pairwise and higher-order interactions, with their first neighbors. Previous work has shown that, in contrast to spatially structured networks such as lattices, ER networks do not promote stability with pairwise interactions alone \cite{calleja-solanas_structured_2022}. ER networks restrict who interacts with whom without introducing any geometric spatial structure, and therefore provide a clean setting to test whether HOIs can stabilize the dynamics away from the well-mixed assumption.

In this setting, we abandon the analytical description of Eq.~\eqref{eq:general} and rely on individual-based simulations on random Erd\"os-R\'enyi (ER) hypergraphs with a fixed mean number of interactions per node and an increasing fraction of HOIs, $\alpha$. 
Moreover, with the aim of analyzing the role of interaction networks alone, we impose again the simple setting of identical physiological rates.
We create ER networks, whose nodes represent single individuals of the different species connected by links and hyperlinks. To simulate local interactions, we fix the mean number of interactions (links plus hyperlinks) per node, $\langle k \rangle$, and consider $\langle k \rangle=10$, $20$, and $40$, with $N=5\cdot10^4$ and $N=10^4$ individuals.
With $\alpha=0$, the network is only composed of pairwise links, while for larger values of $\alpha$ a fraction $\alpha$ of the interactions of each node are, on average, hyperlinks. Because at each update an interaction of the focal node is chosen uniformly at random, $\alpha$ is also the probability that a contest is three-way, as in the well-mixed model. Note, however, that on the network $\alpha$ additionally changes the topology, since it controls how many individuals share hyperlinks (see Appendix~\ref{sec:methods_net} for details). Since, without stabilization, every run eventually loses a species, the probability $p_{\text{coex}}$ that all three species persist is measured over a finite window, which must be long compared with the time it takes an unstable trajectory to reach the boundary (SM Section~S3 and Fig.~S7). For this reason we use a window of $20000$ time units. 

\begin{figure*}[t]
    \centering
    \includegraphics[width=\linewidth]{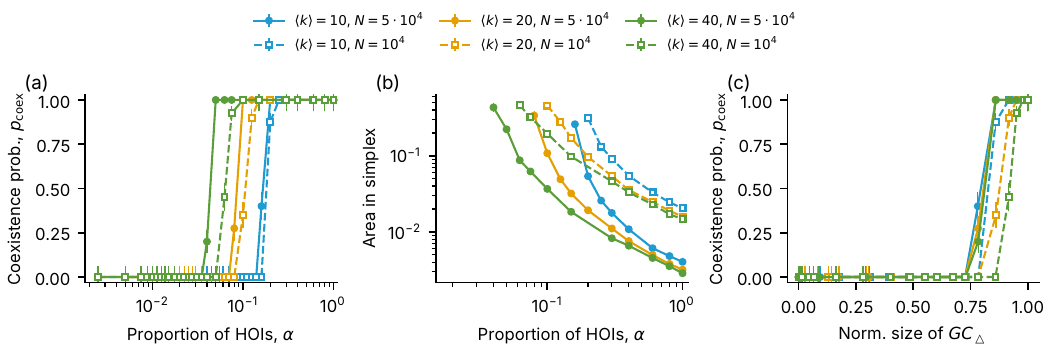} 
    \caption{\textbf{On sparse random interaction networks, HOIs stabilize community dynamics only above a finite fraction, which decreases inversely with the mean degree.} Individual-based simulations on Erdős-Rényi hypergraphs with identical physiological rates, using the dominance matrix introduced in \cite{calleja-solanas_structured_2022}, for mean number of interactions per node $\langle k\rangle=10$ (blue), $20$ (orange), and $40$ (green), with $N=5\cdot10^4$ (filled symbols, solid lines) and $N=10^4$ (open symbols, dashed lines). Each point averages $8$ runs, or $40$ runs near the threshold ($0.8\le\langle k\rangle\alpha\le4$), with a new hypergraph every $4$ runs; each run lasts $20000$ time steps after a transient of $100$; error bars show the binomial standard error of $p_{\textrm{coex}}$ and the standard error of the mean of the area and of $GC_\triangle$. (a) Probability $p_{\textrm{coex}}$ that all species persist until the end of the run. (b) Area in the simplex covered by the trajectory, averaged over runs in which all species persist (shown when at least $3$ runs persist). (c) $p_{\textrm{coex}}$ as a function of the normalized size of the giant component formed by hyperlinks only, $GC_\triangle$.}
\label{fig:5}
\end{figure*}

Figure~\ref{fig:5}a shows the probability that all three species coexist until the end of the simulation as a function of the proportion of HOIs. In contrast with the well-mixed case with the same, identical physiological rates, on sparse random hypergraphs a finite fraction of HOIs is needed. We define the threshold $\alpha^*$ as the value of $\alpha$ at which $p_{\text{coex}}=1/2$. The threshold $\alpha^*$ shifts with the mean degree. As $\langle k\rangle$ grows, approaching the mean-field limit, $\alpha^*$ decreases, however, still maintaining values in the order of $\alpha \simeq 0.1$. Above the threshold, the area in the simplex covered by the trajectory decreases steadily with $\alpha$ (Fig.~\ref{fig:5}b). Smaller populations require more HOIs and display larger oscillations, as expected from stronger demographic fluctuations.


Having determined the minimum amount of HOIs needed for stability, we now ask what controls this threshold. The relevant quantity turns out to be the mean number of hyperlinks in which each node takes part, $\langle k\rangle\alpha$, rather than the fraction of HOIs. For $N=5\cdot10^4$, the threshold corresponds to $\langle k\rangle\alpha^*\approx1.7$ for all three mean degrees: stabilization requires each individual to take part, on average, in more than one HOI. The transition is sharp: for $N=5\cdot10^4$ and all three mean degrees, no run keeps the three species when $\langle k\rangle\alpha\le1.4$, and all runs do when $\langle k\rangle\alpha\ge2$. At this density, hyperlinks percolate through almost the entire network. A giant component formed by hyperlinks only, $GC_\triangle$, appears already at $\langle k\rangle\alpha=1/2$ (SM Section~S3 and Fig.~S6), but at the threshold it contains about $80\%$ of the nodes for all mean degrees, so that the curves collapse when plotted against $GC_\triangle$ (Fig.~\ref{fig:5}c): HOIs stabilize the dynamics only when they reach practically the whole population.

Taken together these results reveal two complementary insights. First, when interactions are restricted to a sparse random network, stable coexistence can be achieved only when a finite proportion of HOIs is present, even for identical physiological rates. Second, what controls stabilization is the number of HOIs in which each individual takes part rather than their proportion. These findings suggest that HOIs alone are unlikely to be responsible for coexistence in real communities. Our results also suggest that models focusing solely on one mechanism may miss crucial dynamics that can promote stability through the interplay of more complex interaction patterns.

\section{Discussion \label{sec:4}}
Understanding the mechanisms that enable species coexistence within ecological communities is critical for biodiversity maintenance, ecosystem functioning, and conservation. Recent theoretical models claim that HOIs create conditions for coexistence. This work was motivated by the need to quantify when HOIs are sufficient to stabilize coexistence in intransitive competitive communities. Here, we demonstrate that the theoretical power of HOIs as a stabilizing mechanism weakens when parameters and models assumptions are constrained more realistically.

In particular, we have analyzed different proportions of pairwise and HOIs under different values of physiological rates, ranging from equivalent to distinct species. 
Our results show that a critical proportion of HOIs can stabilize competitive communities that cannot be stable under pairwise interactions alone only under specific settings. With identical species, any nonzero fraction of HOIs suffices. As soon as the ratios of birth to mortality rates differ among species, pairwise competition becomes actively destabilizing, and a finite fraction of HOIs, $\alpha_c$, is needed to counteract it; for small differences, $\alpha_c$ grows proportionally to the variance of these ratios. When physiological rates become more heterogeneous, the transition towards stability often disappears, both because a feasible equilibrium no longer exists and because, when it exists, HOIs cannot stabilize it. Moreover, close to the transition the stable equilibrium typically coexists with an attracting cycle along the boundary of the simplex, so that robust coexistence requires a fraction of HOIs larger than $\alpha_c$. With empirical physiological rates of plants, species differences are large enough that HOIs stabilize coexistence in only about one in ten randomly assembled triplets. 

We also studied the role of HOIs when individuals interact on a sparse random network. In this case, a finite fraction of HOIs is needed even for identical species. Coexistence requires each individual to take part, on average, in more than one HOI, well beyond the percolation of hyperlinks. This result reinforces the idea that the effect of higher-order interactions on ecological stability depends not only on their presence, but also on the context in which they act~\cite{terry2025impact}.

We believe that our results challenge the stabilizing effect of HOIs. In fact, their effect is weakened or demands additional conditions when we incorporate more ecological complexity. In reality, species do differ in their birth and mortality rates, and this variability plays a significant role in community dynamics.  This dependence on heterogeneity was something that models based on random interactions could not foresee, and it is not a consequence of the degenerate neutral cycles of the equal-rates limit: with heterogeneous rates the transition to stability is generic. In such cases, HOIs guarantee coexistence, but only if pairwise interactions are weak or facilitative \cite{gibbs2024can}.
    
Our study also has several limitations. Although our framework incorporates more ecological realism than previous models, additional features of natural ecological communities could still be included. However, the absence of universal stabilization by HOIs in our setting suggests that their stabilizing role may not persist in more complex models. To ensure sufficient statistical power, we used synthetic dominance matrices $H$, as interaction data is too scarce and difficult to estimate. We also do not attempt to reproduce specific natural communities. Instead, we use empirical physiological rates to explore their implications within our theoretical framework. Accordingly, sampled triplets should be interpreted as representative combinations drawn from realistic distributions of physiological rates, rather than as coexisting species in nature. Despite this limitation, the trends observed in $\alpha_c$ suggest that physiological variability would likely yield similar results among spatially co-occurring communities. Our analysis is also restricted to three species, the minimal setting in which the mechanism of Ref.~\cite{grilli_higher-order_2017} operates. Finally, we constructed the higher-order dominance tensor $B$ by aggregating the pairwise outcomes in $H$, but this tournament-based method can overlook genuine three-species (or higher) effects that are not reducible to pairwise combinations. In principle, one could replace this tournament-derived method with a direct parametrization of higher-order effects without losing analytical traceability. However, obtaining empirical estimates for these modified interactions is currently challenging. 
 
Along this line and building on the need for empirical studies of higher-order effects, Buche et al.~\cite{buche2024multitrophic} offers a compelling case that combines HOIs with network structure to predict species persistence. In their multitrophic experiments, they show that incorporating HOIs in the model description alters the strength and even reverses the sign of per-capita interactions among plants, leading to substantially different predictions of coexistence than non-HOIs models (in agreement with \cite{sundarraman_higher-order_2020,mayfield_higher-order_2017}). Incidentally, these works also support that HOIs in a well-mixed model alone are insufficient: only when embedded within the pairwise interactions and spatial experiments do models achieve reliable predictions of each species’ persistence probability. They thus emphasize our message that higher-order coefficients must go hand in hand with other coexistence mechanisms to faithfully predict coexistence. 

In the spirit of these empirical works and limitations, our results thus suggest two next steps: first, quantifying how the values of the higher-order dominance tensor $B$ are distributed in natural communities \cite{rosas_disentangling_2022, letten_mechanistic_2019} (whose theoretical predictions are suggested in Gibbs et al.~\cite{gibbs_coexistence_2022}), and second, quantifying how the number of interactions per individual and the strength of demographic fluctuations in real communities compare with the regime in which, according to our network results, HOIs can be effective.

While adding realism decreases the stabilizing effect, this does not allow us to categorically dismiss the importance of HOIs. Despite the potential of combining mechanisms to promote coexistence, we still lack the knowledge needed to disentangle HOIs' importance for coexistence in nature. Progress requires coupling empirical data and observations to our models. In this line, the longstanding debate between theoretical and field ecologists over the value of mathematical abstractions reminds us that the law of parsimony favors simpler models when they perform equally well \cite{pomerantz_higher_1981}. Since the mere act of adding higher‐order (or nonlinear) terms boosts explanatory power \cite{aladwani_ecological_2020}, any claimed biological role for a mechanism must be weighed against that baseline. But Occam's razor retaliates: if adding complexity clearly improves explanatory or predictive power, simplicity should be abandoned. Therefore, we must assemble compelling evidence that invoking HOIs as auxiliary mechanisms is indeed the most effective way to explain species coexistence.



\section*{Data availability}
The code implementing the competitive community model to reproduce the main results presented in the paper is available at \url{https://github.com/marcduransala2000/HOIs_dynamics}. The data covering the physiological rates of different plants is available in the supplementary tables of ref. \cite{marba_allometric_2007}.

\begin{acknowledgments}
We thank S. Suweis for fruitful discussions. This work was partially supported by the María de Maeztu project CEX2021-001164-M funded by the  MICIU/AEI/10.13039/501100011033.  S.M. and V.C-S. acknowledge the Spanish State Research Agency through Project No. PID2021-122256NB-C22 funded by MCIN/AEI/10.13039/501100011033/FEDER, UE. S.M. acknowledges financial  support by the Spanish State Research Agency (MICIU/AEI/10.13039/501100011033) and FEDER (UE) under projects COSASTI (PID2024-157493NB-C22). S.M. also acknowledges support from project “CODE – Coupling Opinion Dynamics with Epidemics”, funded under PNRR Mission 4 "Education and Research" - Component C2 - Investment 1.1 - Next Generation EU "Fund for National Research Program and Projects of Significant National Interest" PRIN 2022 PNRR, grant code P2022AKRZ9, CUP B53D23026080001. V.C-S. acknowledges ``Europa Excelencia" 2023 Project No. EUR2023-143472/AEI/10.13039/501100011033 funded by the Spanish State Research Agency and Recovery plan ``NextGenerationEU''. During the revision of this work, the authors used Claude Opus 5.5 (Anthropic), to help revise the text and prepare the code for sharing; the authors checked all code and text.
\end{acknowledgments}

\appendix

\section{Mean-field equations from the individual-based process \label{sec:methods_mf}}
We consider a well-mixed population of $N$ individuals in which every site is always occupied. At each event, an individual dies; it belongs to species $i$ with probability $d_ix_i/D$. The vacant site is recolonized by the winner of a contest among seeds. With probability $1-\alpha$ the contest involves two seeds and with probability $\alpha$ three seeds, each drawn independently from species $j$ with probability $f_jx_j/F$, proportional to its abundance and birth rate. In a two-seed contest, species $i$ wins with probability $\sum_j 2\,(f_ix_i/F)(f_jx_j/F)\,H_{ij}$, where the factor $2$ accounts for the two possible orderings of the seeds (for $j=i$, $2H_{ii}=1$). In a three-seed contest, $i$ can occupy any of the three positions and wins with probability $B_{ijk}/3$, giving $\sum_{j,k}(f_ix_i/F)(f_jx_j/F)(f_kx_k/F)\,B_{ijk}$. Measuring time so that deaths occur at total rate $N D$, the expected change of $x_i$ per unit time is the probability that species $i$ wins the vacant site, multiplied by $D$, minus its death rate $d_ix_i$. This yields Eq.~\eqref{eq:general}, whose limits $\alpha=0$ and $\alpha=1$ are Eqs.~\eqref{eq:pair} and~\eqref{eq:ho}. For $N\to\infty$ fluctuations vanish, and the Monte Carlo simulations of Fig.~\ref{fig:2}d sample this process at finite $N$.

\section{Lyapunov function for different physiological rates \label{sec:methods_lya}}

We consider the well-mixed dynamics of Eq.~\eqref{eq:general} and use the notation of Eq.~\eqref{eq:T}, so that
\begin{equation}
    \frac{dx_i}{dt} = x_i\left[(1-\alpha)\,T_{1,i} + \alpha\, T_{2,i} - d_i\right].
    \label{eq:app_general}
\end{equation}
Let $\Vec{x}^*$ be a feasible equilibrium for a given $\alpha$, i.e. $x^*_i>0$ and $(1-\alpha)T^*_{1,i}+\alpha T^*_{2,i}=d_i$ for all $i$ [Eq.~\eqref{eq:equilibrium_diff}]. We denote $F^*=F(\Vec{x}^*)$, $D^*=D(\Vec{x}^*)$, and $T^*_{\cdot,i}=T_{\cdot,i}(\Vec{x}^*)$, while $F$, $D$, and $T_{\cdot,i}$ without a star are evaluated at the current state $\Vec{x}$. Note that, unlike the case of equal rates, $\Vec{x}^*$ depends on $\alpha$ and is in general different from the equilibria of the purely pairwise or purely higher-order dynamics. The time derivative of the function $V(\Vec{x})=-\sum_i x^*_i\log(x_i/x^*_i)$ [Eq.~\eqref{eq:Lyapunov}] along trajectories is
\begin{equation}
    \frac{dV}{dt} = -\sum_i \frac{x^*_i}{x_i}\frac{dx_i}{dt}
    = D^* -(1-\alpha)\sum_i x^*_i T_{1,i} - \alpha \sum_i x^*_i T_{2,i},
    \label{eq:app_V1}
\end{equation}
where we used $\sum_i x^*_i d_i=D^*$. Using $H_{ij}=1-H_{ji}$, the pairwise term can be rewritten as
\begin{equation}
\begin{split}
    \sum_i x^*_i T_{1,i} &= \frac{2D}{F^2}\sum_{i,j} f_i f_j H_{ij}\, x^*_i x_j \\
    &= \frac{2D}{F^2}\left(F^*F - \sum_{j} f_j x_j\sum_i H_{ji} f_i x^*_i \right) \\
    &= \frac{2DF^*}{F} - \frac{F^{*2} D}{F^2 D^*} \sum_j x_j T^*_{1,j},
\end{split}
\end{equation}
where in the last step we recognized the expression of $T^*_{1,j}$. Substituting into Eq.~\eqref{eq:app_V1} and using the equilibrium condition in the form $(1-\alpha)\sum_j x_j T^*_{1,j} = D - \alpha\sum_j x_j T^*_{2,j}$, we obtain
\begin{equation}
\begin{split}
    \frac{dV}{dt} &= D^* - \frac{2DF^*}{F} + \frac{F^{*2}D^2}{F^2D^*} + \alpha\, G(\Vec{x}) \\
    &= \frac{\left(F^*D - FD^*\right)^2}{F^2D^*} + \alpha\, G(\Vec{x}),
\end{split}
\label{eq:app_dV}
\end{equation}
with
\begin{equation}
    G(\Vec{x}) = \frac{2DF^*}{F} - \sum_i x^*_i T_{2,i} - \frac{F^{*2}D}{F^2D^*}\sum_i x_i T^*_{2,i}.
\end{equation}
We verified Eq.~\eqref{eq:app_dV} numerically against the direct evaluation of Eq.~\eqref{eq:app_V1} for random states, physiological rates, and dominance matrices. For equal physiological rates ($f_i=d_i=1$), $F=D=1$, the first term vanishes, and Eq.~\eqref{eq:app_dV} reduces to Eq.~\eqref{eq:Lyapunov_derivation1}. The first term vanishes identically also when $f_i=c\,d_i$ for all $i$, since then $F=cD$ for every $\Vec{x}$.


We also assessed the basin of attraction of the stable equilibrium by integrating Eq.~\eqref{eq:general} (LSODA) from $20$ initial conditions drawn uniformly in the simplex, up to $t=10^5$. A trajectory was classified as converging if it ended within $10^{-3}$ of the equilibrium, and as reaching the boundary if any density fell below $10^{-8}$, which corresponds to extinction in any finite population.

\section{Linear stability analysis and numerical procedure \label{sec:methods_jac}}
\paragraph*{Jacobian.}
To assess the local stability of feasible interior equilibria of Eq.~\eqref{eq:general}, we write
\begin{equation}
    \frac{dx_i}{dt}=x_i R_i,
    \qquad R_i=(1-\alpha)\,T_{1,i}+\alpha\, T_{2,i}-d_i,
\end{equation}
where $T_{1,i}$ and $T_{2,i}$ are defined in Eq.~\eqref{eq:T}. Using the definitions of $F$ and $D$, together with $B_{ijk}=B_{ikj}$, differentiation with respect to $x_m$ gives
\begin{align}
    \partial_m T_{1,i} &= T_{1,i}\left(\frac{d_m}{D}-\frac{2f_m}{F}\right)+\frac{2Df_if_mH_{im}}{F^2},
    \label{eq:jac_dT1}\\
    \partial_m T_{2,i} &= T_{2,i}\left(\frac{d_m}{D}-\frac{3f_m}{F}\right)+\frac{2Df_if_m}{F^3}\sum_k B_{imk}f_kx_k,
    \label{eq:jac_dT2}
\end{align}
where $\partial_m=\partial/\partial x_m$. Defining $M_{im}=\partial_m R_i=(1-\alpha)\,\partial_m T_{1,i}+\alpha\,\partial_m T_{2,i}$, the Jacobian is
\begin{equation}
    \mathcal{J}_{im}=\frac{\partial\dot{x}_i}{\partial x_m}
    =\delta_{im}R_i+x_iM_{im},
    \label{eq:jac_full}
\end{equation}
where $\delta_{im}$ is the Kronecker delta. At a feasible interior equilibrium, $x_i^*>0$, $\sum_i x_i^*=1$, and $R_i^*=0$, so that $\mathcal{J}_{im}^*=x_i^*M_{im}^*$.

Since the dynamics conserve total density, only $g-1$ species densities are independent. We therefore restrict the Jacobian to the simplex by eliminating $x_g=1-\sum_{a=1}^{g-1}x_a$. The resulting reduced Jacobian is
\begin{equation}
    J_{ab}=\mathcal{J}_{ab}^*-\mathcal{J}_{ag}^*=x_a^*(M_{ab}^*-M_{ag}^*),
    \qquad a,b=1,\ldots,g-1.
    \label{eq:jac_reduced}
\end{equation}
Because $\sum_i dx_i/dt=0$ for every $\Vec{x}$, the columns of $\mathcal{J}^*$ sum to zero; $\mathcal{J}^*$ thus has a zero eigenvalue, associated with changes of the total density, and its other eigenvalues are those of $J$, so that in particular $\mathrm{tr}\,J=\mathrm{tr}\,\mathcal{J}^*$. An interior equilibrium is locally asymptotically stable if all eigenvalues of $J$ have strictly negative real parts, and unstable if at least one eigenvalue has positive real part.

\paragraph*{Only pairwise interactions.}
For $\alpha=0$, the equilibrium condition gives $T_{1,i}^*=d_i$. Using $H_{ii}=1/2$ and Eq.~\eqref{eq:jac_dT1}, we obtain
\begin{equation}
\begin{split}
    \mathrm{tr}\,J(0) &= \sum_i x_i^*\left(\frac{d_i^2}{D^*}-\frac{2d_if_i}{F^*}+\frac{D^*f_i^2}{F^{*2}}\right)\\
    &=\frac{\sum_i x_i^*\left(F^*d_i-D^*f_i\right)^2}{F^{*2}D^*}.
\end{split}
    \label{eq:jac_pairwise_trace}
\end{equation}
For a feasible interior equilibrium, this expression is strictly positive whenever the ratios $f_i/d_i$ are not all identical. Thus, at least one eigenvalue has positive real part, and such pairwise equilibria are unstable.

\paragraph*{Equal physiological rates.}
For equal physiological rates ($d_i=f_i=1$), any feasible interior equilibrium of the pairwise model ($2\sum_j H_{ij}x_j^*=1$) is also an equilibrium when HOIs are introduced, independently of their proportion $\alpha$. To study its local stability, we define $A_{ij}=H_{ij}-1/2$ and $P=\mathrm{diag}(x_1^*,\ldots,x_g^*)$. Since $H_{ij}+H_{ji}=1$, we have $A^{\mathsf T}=-A$ and $\sum_j A_{ij}x_j^*=0$. The Jacobian matrices evaluated at the equilibrium for purely pairwise and purely higher-order interactions are
\begin{equation}
    \mathcal{J}_{\mathrm{pair}}^*=2PA,
    \qquad
    \mathcal{J}_{\mathrm{HOI}}^*=3PA+2PAPA.
\end{equation}
For an arbitrary proportion of HOIs, the reduced Jacobian therefore satisfies
\begin{equation}
    J(\alpha)=\left(1+\frac{\alpha}{2}\right)J(0)+\frac{\alpha}{2}J(0)^2,
    \label{eq:jac_Kalpha}
\end{equation}
where $J(0)$ corresponds to purely pairwise interactions.

\paragraph*{Three-species communities.}
For the three-species communities considered here ($g=3$), $J$ is a $2\times2$ matrix. Defining $\tau=\mathrm{tr}\,J$ and $\Delta=\det J$, the linearized system is asymptotically stable if and only if
\begin{equation}
    \tau<0,\qquad \Delta>0.
    \label{eq:jac_stability}
\end{equation}
For equal physiological rates, assuming a feasible and isolated interior equilibrium, the eigenvalues of $J(0)$ are $\pm i\omega_0$, where
\begin{equation}
    \omega_0^2=4\sum_{i<j}x_i^*x_j^*A_{ij}^2>0.
\end{equation}
Using Eq.~\eqref{eq:jac_Kalpha}, the eigenvalues for any $\alpha$ are
\begin{equation}
    \lambda_{\pm}(\alpha)=-\frac{\alpha\,\omega_0^2}{2}\pm i\left(1+\frac{\alpha}{2}\right)\omega_0.
    \label{eq:jac_eig}
\end{equation}
Thus, the equilibrium is locally asymptotically stable for any $\alpha>0$, while for $\alpha=0$ the eigenvalues are purely imaginary, consistent with the neutral oscillations obtained from the Lyapunov analysis (Fig.~\ref{fig:2}c).

\paragraph*{Different physiological rates.}
The closed-form results above rely on the fact that, for equal rates, the equilibrium does not depend on $\alpha$ and is known explicitly. For different physiological rates this is no longer the case. The equilibrium $\Vec{x}^*(\alpha)$ is a solution of the nonlinear equations~\eqref{eq:equilibrium_diff}, which in general cannot be solved in closed form, and it moves with $\alpha$, so that the entries of $J$, and hence $\tau$ and $\Delta$, are no longer simple functions of $\alpha$. In addition, a feasible equilibrium may exist only for some values of $\alpha$, and more than one may exist for the same $\alpha$. Apart from the instability of the purely pairwise equilibrium, Eq.~\eqref{eq:jac_pairwise_trace}, the analytical approach therefore does not provide the threshold $\alpha_c$, and neither does the Lyapunov function, whose derivative is not sign-definite for $\alpha>0$ (Appendix~\ref{sec:methods_lya}). We therefore locate the feasible equilibria numerically for each value of $\alpha$ and assess their local stability through the eigenvalues of the reduced Jacobian, Eqs.~\eqref{eq:jac_reduced} and~\eqref{eq:jac_stability}, as described below.

\paragraph*{Numerical procedure.}
Using $x_3=1-x_1-x_2$, we reduce Eq.~\eqref{eq:general} with $g=3$ to a two-dimensional system for $(x_1,x_2)$. Since multiplying all birth and mortality rates by a common factor only rescales time, we normalize all rates by their maximum value within each triplet of species. For each value of $\alpha$ on a grid of $81$ equally spaced values in $[0,1]$, we search for equilibria by solving the reduced equilibrium equations with a Newton-type root finder (MINPACK \texttt{hybrd}, as implemented in \texttt{scipy.optimize.fsolve}) started from $12$ initial conditions on a regular grid in the interior of the simplex and from the equilibria found at the previous value of $\alpha$ on the grid. We retain solutions with residual smaller than $10^{-10}$ and $x_i>10^{-9}$ for all $i$ (feasibility), and merge duplicates. At each feasible equilibrium we compute the $2\times2$ Jacobian by central finite differences and classify the equilibrium as locally asymptotically stable if both eigenvalues have negative real part. The critical value $\alpha_c$ is the smallest $\alpha$ on the grid with at least one feasible and stable equilibrium.

To validate this procedure, we compared it with the criterion based on direct integration of Eq.~\eqref{eq:general} with a fourth-order Runge–Kutta scheme, in which a transition is detected when the amplitude of the oscillations decreases over time. For $300$ triplets of normally distributed rates ($\sigma$ between $0.01$ and $0.5$), the two methods agreed in all cases on the existence of a transition, and the mean absolute difference in $\alpha_c$ ranged from $0.004$ to $0.03$, with the linear stability estimate being slightly lower, as expected since the integration-based criterion requires a finite decay of the oscillations to be observed within the integration time.

\section{Competitive community model on interaction networks \label{sec:methods_net}}
To relax the well-mixed assumption in our model we extend it to consider interactions taking place on a network with an increasing fraction of higher-order interactions. 

We place each individual in a node, which symbolizes a fixed spatial location (Fig.~\ref{fig:1}). Individuals compete to place their offspring in an empty node only if there is a link between them. These links can now be of two different types based on the number of individuals involved (i.e., the interaction order). Pairwise, connecting only two individuals, and higher-order links with three individuals at time. 

The initial configuration is set by randomly distributing the species among the $N$ nodes. We then build an Erdős-Rényi hypergraph by adding pairwise links between each pair of nodes with probability $p_1$ and hyperlinks among each triplet of nodes with probability $p_2$. We choose $p_1=\langle k\rangle(1-\alpha)/(N-1)$ and $p_2=2\langle k\rangle\alpha/[(N-1)(N-2)]$, so that each node has on average $\langle k\rangle(1-\alpha)$ pairwise links and $\langle k\rangle\alpha$ hyperlinks, i.e. a mean total number of interactions $\langle k\rangle$ independent of $\alpha$, of which a fraction $\alpha$ are higher-order. Note that, with this construction, the forgetful degree~\cite{battiston_networks_2020}, in which each hyperlink counts twice, is $\langle k\rangle(1+\alpha)$ on average. We keep the largest connected component of the resulting hypergraph, which in all our simulations contains more than $99\%$ of the nodes, and compute $GC_\triangle$ as the largest connected component of the hypergraph formed by hyperlinks only, normalized by the number of nodes.

At each elementary update, a node is chosen at random, and one of its interactions (a link or a hyperlink) is chosen uniformly at random. The species occupying the node is then replaced by the winner of a contest among the individuals involved in that interaction: in a pairwise contest the outcome is decided by $H$, and in a three-way contest species $i$ wins according to $B_{ijk}$. A unit of time corresponds to $N$ elementary updates. Each run lasts $20000$ time units after a transient of $100$, and stops if a species goes extinct. The area in the simplex is computed as described in SM Section~S1. The simulations were performed with a compiled reimplementation of the code in the repository listed below, which follows the same update rules. 

\bibliography{references}

\end{document}